\documentstyle[aps]{revtex}

\begin{document}

\draft
\preprint{AZPH-TH/94-8}
\title{
Thermally Induced Density Perturbations in the Inflation Era}
\author{Arjun Berera\cite{byline} and Li-Zhi Fang}
\address{Department of Physics, University of Arizona, Tucson, 85721}
\date{\today}
\maketitle

\begin{abstract}
The possibility of thermally induced initial density perturbations in
inflationary cosmology is examined.
The fluctuation dynamics of a scalar
field plus thermal bath system during slow roll is described
by a Langevin-like equation.
Fluctuation-dissipation arguments
show that for a wide parameter range within the standard inflation model,
the thermal fluctuations of the scalar field
can dominate its quantum fluctuations.
The initial amplitude of density perturbations is found to
lie in a range which is consistent with the recent observations of cosmic
temperature fluctuations.
\end{abstract}

\narrowtext

The standard scenario for structure formation in the universe is
based on inflationary cosmology. According to this model,
quantum fluctuations of the scalar field during the expansion
era were the perturbing seeds in an initial, globally
smooth universe. From this, large-scale structures then arose \cite{kolb}.
This model predicts that the initial density perturbations should
be Gaussian, and have a power-law spectrum with index $n \sim 1$.
Based on this model for the initial perturbations plus assumptions
about dark matter, the formation of galaxies, clusters of galaxies and
other objects have been extensively studied \cite{peebles}.
Subsequent studies have shown that this model is consistent with
observation, thus not theoretically implausible. Recent detection of
temperature anisotropies in cosmic background radiation (CBR) by the
Differential Microwave Radiometer (DMR) on the Cosmic Background
Explorer's ({\it COBE}) satellite has given the first opportunity
to directly probe the initial density perturbation. These results fit
the scaling spectrum given by the inflation model \cite{smoot}. Thus
it is one source of support for the inflation model in describing
the initial density perturbations.

However, the {\it COBE}-DMR results do raise questions about the
amplitude of the initial perturbations. Such questions are not new,
only further perpetuated by the {\it COBE} results. Before {\it COBE}
it was already possible to determine the amplitude by fitting (or
normalizing) the evolved perturbations with the observed clustering
of galaxies on scales of say 8 h$^{-1}$Mpc, where $h$ is the Hubble
constant in unit of 100 km s$^{-1}$Mpc$^{-1}$.
This method suffers from three uncertainties arising
from:  1) the evolution of the perturbations; 2) the assumptions of
dark matter, and 3) the bias factor. On the other hand, the temperature
fluctuations of CBR on scales of superhorizons size at the decoupling
time, i.e. larger than about 2$^{\circ}$, directly provide information
about the initial density perturbation, independent of the above three
factors. The {\it COBE} result of the CBR quadrupole amplitude has
already been used to normalize the amplitude of the cosmic density
perturbations.  Therefore, it is now a question of observational
importance to explain the so found amplitude of the initial perturbations.

The origin of the amplitude's magnitude is also relevant because the
"standard" scenario cannot explain it naturally. It has been known
since developing the inflation model, that the amplitude of the initial
density perturbations given by quantum fluctuations of the inflationary
scalar field is, at least for the "standard" model, of order 1, which is
about 10$^4$  times larger than the required value
\cite{brand}. This is sometimes called the fluctuation problem.  One can
relax this conflict by sophisticated designs for the inflationary potential.
However, such particle physics designs counter the naturality philosophy of
inflation. Moreover, in these models the amplitude is completely
determined by unknown parameter(s) of the so designed potential. This gives
almost no constraint to the possible or reasonable range for the
amplitude. As such, it cannot predict what the initial value of the
perturbation amplitude should be.

This situation motivates a search other possible mechanisms, which
do not depend on such sophisticated design. In this paper we show how
thermal fluctuations during inflation may actually play the dominant
role in producing the initial perturbations. It is conventionally
believed that the components of particle-like matter, either relativistic
or non-relativistic, are totally negligible during inflation. This
is certainly true if we only consider the energy density, because
inflation is by definition the epoch when the vacuum energy of the
scalar field was the dominant component in the universe. However, this
does not imply that the initial perturbations must mainly arise
from quantum fluctuation of the scalar field. All particle-like matter
which existed before inflation would have been dispersed by inflation.
Yet, particle-like matter will not completely vanish if one
considers the processes of the scalar field
dissipating into a thermal bath
via its interaction with other fields.

The existence of a thermal component during inflation may not be
exceptional and perhaps even inevitable.
Roughly one can see this as follows. In order to
maintain the $\phi$
field close to its minimum at the onset of the inflation phase transition,
thermal forces will generically be an important contributing source.
Therefore, at least in the starting period of the phase transition,
there is thermal contact between $\phi$ and all other fields
with which it interacts. During the slow roll period
of inflation, the kinetic and potential energy of the $\phi$ field
is fairly constant, so
that the interaction between the $\phi$ field with the other fields
remains about the same as at the beginning. As such, there is
no compelling reason to believe that the thermal component
vanishes during inflation.

We will show that, for a wide range of the parameters characterizing
inflation, there can be a thermal component in the universe.
With account for such a component, we find
that the density perturbations can originate mainly from thermal
fluctuations. Accepting this mechanism, the fluctuation problem would
be automatically avoided. In addition the allowable range for the
amplitude would be consistent with the observed results.

Let us consider the standard model of inflation, given by a scalar
field $\phi$ with Lagrangian density
\begin{equation}
L={\frac{1}{2}}\partial^{\mu}\phi\partial_{\mu}\phi- V(\phi) + L_{int}
\ ,
\end{equation}
where $V(\phi)$ is the effective potential, and
$L_{int}$ describes the interaction of $\phi$ with all other fields.
The classical equation of motion for $\phi$ in a de-Sitter universe is
\begin{equation}
\ddot{\phi} + 3H \dot{\phi} + \Gamma_{\phi}\dot{\phi}
- e^{-2Ht}\nabla^2\phi+ V'(\phi)=0 \ .
\end{equation}
The friction term $\Gamma_{\phi}\dot{\phi}$ phenomenologically describes
the decay of the $\phi$ field via the interaction Lagrangian $L_{int}$.
In principle, the friction term $\Gamma_{\phi} \dot{\phi}$ may not be
reasonable for describing the energy transfer
from $\phi$-field particle production during far out of equilibrium
conditions. However, it is a proper approximation for describing the energy
dissipated by the $\phi$ field into a thermalized radiation bath.
It should also be noted that $\Gamma_{\phi}$ could be a function of
$\phi$. Due to the lack of a detailed model for the decay of $\phi$, we
assume that $\Gamma_{\phi}$ is a constant independent of $\phi$.
As we will show, the thermal fluctuations of $\phi$ do not
depend significantly on the details of $\Gamma_{\phi}$.

In the standard treatment of the inflation model, one assumes
that the inflation era is divided into two regimes: 1) slow roll
and 2) reheating. In the former all interactions between the inflationary
scalar field and other fields are typically neglected.
These interactions have been considered only in the latter,
in order to supply the mechanism
to reheat the universe. This is equivalent to assuming that the friction
term is only important in the reheating regime, but negligible during the
slow-roll (inflation) regime, i.e. $3H \gg \Gamma_{\phi}$.
In this case, the slow-roll evolution in a spatially homogeneous
universe is given by
\begin{equation}
\dot{\phi} \simeq - \frac {V'(\phi)}{3H}
\ ,
\end{equation}
where $H^2=(8\pi G/3)\rho_{\phi} \sim (8\pi/3)M^4/m_{pl}^2$,
$\rho_{\phi}=\dot{\phi}^2/2 + V(\phi)$ is energy of the $\phi$ field,
$m_{pl} = (1/G)^{1/2}$ is the Planck mass, and $V(\phi) \sim V(0)=M^4$.
Eq.(3) is valid when the potential $V(\phi)$ satisfies
the following well known conditions for the inflationary potential
\cite{kolb},
\begin{center}
$|V^{''}(\phi)| \ \ll \ 24\pi V(\phi)/m^2_{pl}$,
\end{center}
\begin{equation}
V'^2(\phi)m_{pl}^2 \ \ll \  48\pi V^2(\phi)
\ .
\end{equation}

Our first observation is that the condition $\Gamma_{\phi} \ll 3H$ is
not necessary for a slow roll solution.  The coupling of the
inflationary scalar field with other fields can co-exist with the roll
down solution.
Consider the case when $\Gamma_{\phi}$ is comparable to $H$.  This implies
that its decay products will equilibrate quickly to
some temperature $T_r$.  For explicitness in the treatment below, let
us make the reasonable assumption that the decay products
of the $\phi$-field are relativistic matter.
The additional equation needed to describe this relativistic component
from the first law of thermodynamics is
\begin{equation}
\dot{\rho}_{r} + 4H\rho_{r}=\Gamma_{\phi}\dot{\phi}^2
\ ,
\end{equation}
where $\rho_{r}$ is the energy density of the thermal component.
With account for this component, the slow roll equation (3) should
be replaced by the set
\begin{equation}
\dot{\phi} \simeq - \frac {V'(\phi)}{3H + \Gamma_{\phi}}
\end{equation}
\begin{equation}
\dot{\rho}_r \simeq 0.
\end{equation}
and
\begin{equation}
H^2={\frac{8\pi G}{3}}(\rho_{\phi} + \rho_{r})
\ .
\end{equation}
Strictly speaking, when there is a thermal component in the universe,
one should use a finite temperature effective potential to replace
the zero-temperature potential $V(\phi)$.
As we will show below, this replacement will not be important for
our purpose.

Solution (7) implies that independent of the initial conditions
for the thermal component, it will
reach a steady state regime during inflation,
i. e. the depletion of the radiation due to expansion will be balanced by
its production due to friction. From eqs.(5) - (7), the constant
energy density of the thermal component is found to be
\begin{equation}
\rho_r \simeq \frac{\Gamma_{\phi}}{4H} \dot{\phi}^2
\ .
\end{equation}
In the inflation epoch the kinetic energy of the $\phi$ field,
$\dot \phi^2/2$, is much less than its vacuum energy
$\rho_{\phi} \sim V(\phi)$.
Thus we have $\rho_r \ll \rho_{\phi}$ if
\begin{equation}
\Gamma_{\phi} \leq \alpha 4H
\ ,
\end{equation}
where $\alpha > 1$ is a model dependent arbitrariness.
As such, in terms of energy density, the considered system during
inflation is still dominated by the vacuum energy of the $\phi$ field,
with in particular the thermal component being negligible.  Thus
other aspects of the
the inflationary scenario will remain the same
as in the standard model.

However, in terms of the system's temperature, the presence
of a thermal component is not necessarily negligible. The temperature,
$T_r$, of the thermal component is given by
\begin{equation}
T_r \simeq \rho_r^{1/4} \simeq (m_{pl}W\Gamma_{\phi})^{1/4} M^{1/2}
\ ,
\end{equation}
where $W=\dot{\phi}^2/2V(\phi)$ is the ratio of the kinetic and
potential energy of the $\phi$ field. From eqs. (9), (10) and (11), it is
easy to show that the temperature of the thermal component can be
greater than the Hawking temperature, i.e.
\begin{equation}
T_r > H
\end{equation}
if
\begin{equation}
\Gamma_{\phi} > \left(\frac{M}{m_{pl}}\right)^5\frac{M}{W} \ .
\end{equation}

All conditions (4), (10) and (13) can be simultaneously satisfied if
\begin{equation}
V^{3/2}(\phi)m^{-3}_{pl} \ll V'(\phi) \ll m_{pl}^{-1} V(\phi)
\end{equation}
Since $(M/m_{pl})^2 \ll 1$, this condition can
be fulfilled.
In fact, there is big room in the parameter space of the potential,
in which the inflationary expansion of the universe will still be
dominated by the scalar field, but the temperature of the system
will be determined by the thermal matter $\rho_r$.
On the other hand,
because $W \ll 1$ during the slow-roll regime,
eq.(11) implies that $T_r \ll M$.  Using $\lambda \phi^4$
as a generic inflationary potential, the leading
temperature effect is known to be
$\lambda T_{r}^2$ \cite{kolb}.  Recalling that
to have inflation requires
$\lambda \ll (M/m_{pl})^2$
means $\lambda T_r^2 \ll M^4/m_{pl}^2
\sim H^2$. Therefore,
the influence of the finite temperature effective potential
is insignificant as stated earlier.

The quantum mechanical fluctuations of the $\phi$ field during
inflation are determined by the Hawking temperature $H$. Therefore,
one can expect that eq.(12) is the condition under which thermal
fluctuations will compete with quantum fluctuation. We will
justify this point in the following.

To calculate the fluctuations, the
$\phi$ field should be treated as a stochastic field. Therefore,
eq.(6) should be interpreted as the ensemble averaged equation of
motion.  The essence of eq.(6) can easily be seen if it is
rewritten as
\begin{equation}
\frac {d\phi}{dt} =  - \frac{1}{3H + \Gamma_{\phi}}
\frac{dF[\phi]}{d \phi}
\ , \end{equation}
where $F[\phi] = V(\phi)$.
Eq.(15) is, in fact, a equation for the rate of change of the order
parameter $\phi$ of a homogeneous system with free energy $F[\phi]$.
It describes the approach to equilibrium for the system.
More generally in a study of fluctuations, one should not use the
approximation of a spatially homogeneous universe, so that
the spatial gradient term  $\exp(-2Ht)\nabla^2\phi$ in eq.(2)
should not be ignored. The equation (15) for the rate of
change is then modified to
\begin{equation}
\frac {d\phi({\bf x}, t)}{dt} =  - \frac{1}{3H + \Gamma_{\phi}}
\frac{\delta F[\phi({\bf x}, t)]}{\delta \phi}
\end{equation}
where the free energy is given by
\begin{equation}
F[\phi]=\int d^3{\bf x} \left[ \frac{1}{2} (e^{-Ht}\nabla\phi)^2 +
V(\phi) \right]
\ .
\end{equation}
Our present purpose is limited to comparing the amplitudes of thermal and
quantum fluctuations.  For this, it is enough only to calculate the
$\phi$-field fluctuations for the mode of wavelength equal to the horizon
$H^{-1}$ during inflation. As such the contribution
from the spatial gradient term for only this mode is important
in calculating the correlation function of the scalar field.

It is known that rate equations like (17) cannot correctly describe
the approach to equilibrium during a phase transition without also a
noise term \cite{gold}. For instance, eq.(17) will only cause the order
parameter
to evolve towards local minima but not necessarily the global minimum.
To ensure that the system approaches the global minimum, we must remember
that actually the order parameter dynamics is not purely relaxational,
but may exhibit fluctuations, arising from the microscopic degrees of
freedom. These fluctuations can be modeled by introducing a noise term
$\eta$ into eq.(17) as
\begin{equation}
\frac {d\phi}{dt} =  - \frac{1}{3H + \Gamma_{\phi}}
\frac{\delta F}{\delta \phi} + \eta(t)
\ .
\end{equation}
This is the Langevin equation for a system with one degree of freedom,
in similar analogy to say a Brownian particle.
A similar type of equation has been examined in \cite{star}.
However, their purpose was to statistically treat the quantum fluctuations
of the scalar field, whereas ours is to treat external thermal forces.

We had shown
in \cite{berera} that during the eras of dissipations,
the dynamics of structure
formation in the universe should be described by a KPZ-equation
\cite{kardar}, which describes systems governed by non-linear effects
plus stochastic fluctuations.
Eq.(18) is a realization of this hypothesis,
within the context of scalar field dynamics in the standard
inflationary model.
Although it is in all generality non-linear, in this paper we will
not study the complete
spectrum of the density perturbations, but only the amplitude
of fluctuations with wavelength $\sim H^{-1}$. For this we will not
concentrate on the non-linear effects in eq.(18).

The stochastic force $\eta$ in eq.(18) can be found from the
fluctuation-dissipation
theorem\cite{hohe}. If the temperature of the thermal bath is
$T_r$, the expectation values of $\eta$ is given by
\begin{equation}
\langle \eta(t)\rangle_{\eta} = 0
\end{equation}
and
\begin{equation}
\langle \eta(t) \eta(t') \rangle _{\eta} = D\delta(t-t')
\ .
\end{equation}
The notation $\langle ...\rangle_{\eta}$ denotes averaging of
$\eta$ with respect to a
Gaussian distribution.  The variance $D$ is given by
\begin{equation}
D=2\frac{1}{U} \frac{T_r}{3H+\Gamma_{\phi}}
\end{equation}
where $U = (4\pi/3) H^{-3}$ is the volume with Hubble radius
$H^{-1}$.

The fluctuations $\delta \phi$ of the $\phi$ field can be found
from linearizing eq.(18). If we only consider the fluctuations
$\delta \phi$ crossing outside the horizon, i.e. with wavelength
$\sim H^{-1}$. The equation of $\delta \phi$ is
\begin{equation}
\frac{d \delta \phi}{dt}= -
\frac{H^2 + V^{''}(\phi)}{3H+\Gamma_{\phi}}\delta \phi + \eta.
\end{equation}
According to condition (4), the term $V^{''}(\phi)$ on the right hand
side of eq.(22) can be neglected. From eq. (22), one finds for
the correlation function of the fluctuations
\begin{equation}
\langle \delta \phi(t) \delta \phi(t') \rangle_{\eta}
\simeq D \frac{3H+\Gamma_{\phi}}{2H^2} e^{-(t-t')H^2/(3H +\Gamma_{\phi})},
\ \ \ t > t',
\end{equation}
so that
\begin{equation}
\langle (\delta \phi)^2 \rangle_{\eta}
\sim \frac{3}{4\pi}HT_r
\ .
\end{equation}
This is our central result.  Notice that it is independent of
$\Gamma_{\phi}$.  This is expected since it is simply the variance
of the $\phi$ field when coupled to a thermal bath, as implicit to the
fluctuation-dissipation theorem.  Nevertheless, the importance of a
sufficiently large decay term, as emphasized
earlier, is to ensure appropriate dynamical conditions for
rapid thermalization of the radiation bath
on the scale of the expansion rate $H$. From eq.(24) one can
conclude that the thermal fluctuations of the scalar
field will be greater than its quantum fluctuations,
$\langle (\delta \phi)^2 \rangle_{QM} \sim H^2/2\pi$, when
condition (12) holds.

Since the kinematics and dynamics of inflation are the same here
as in the standard model, the initial perturbations will still have a
power-law spectrum with index $n \sim 1$. Also the amplitude
$\delta \rho/\rho$, when it crosses back inside the horizon, can
be calculated by the gauge invariant amplitude
$\zeta =\delta \rho_{\phi}/(\rho + p)$ during the time of inflation.
For quantum fluctuations it is known that $\zeta$ is of order 1.
This follows simply because for a field in its ground state, the mean
quantum
fluctuation of its energy density is of the same order as its
mean kinetic energy density.
However, for thermal fluctuations,
the mean kinetic energy density will be greater than its
fluctuation, so one can except that the quantity $\zeta$
should be much less than 1.

The energy fluctuations caused by $\delta \phi$ is $\delta \rho =
\delta \phi V'(\phi)$, and $\rho + p = \rho_{\phi} + p_{\phi} + \rho_r + p_r
= \dot {\phi} ^2 + (4/3) \rho_r$. Therefore from eqs.(6), (9), (11)
and (24), we have
\begin{equation}
\frac{\delta \rho}{\rho}
\simeq \frac{\delta \phi V'(\phi)}{\dot{\phi}^2+(4/3)\rho_r}
\simeq \frac{3}{4} \left(\frac{3\Gamma_{\phi}}{\pi H}\right)^{1/2}
\left(\frac {H}{T_r} \right)^{3/2}
\end{equation}
Because $T_r < M$, eq.(25) shows that the amplitude of the initial
perturbations, $\delta \rho/\rho$, should be in the following range,
\begin{equation}
\left(\frac {M}{m_{pl}} \right)^{3/2} \ll \frac{\delta \rho}{\rho}
\left(\frac{H}{\Gamma_{\phi}}\right)^{1/2}
\ll  1
\ .
\end{equation}
Therefore, the amplitude of the initial perturbations is mainly limited
by the ratio $M/m_{pl}$, i.e. the energy scale of inflation.
If we take $M\sim 10^{15} Gev$, the possible range for the amplitude
$\delta \rho/\rho$
should be in about the middle of the range
$(10^{-6} - 1)(\Gamma_{\phi}/H)^{1/2}$.
This result is consistent with the observed amplitude
$\delta \rho/\rho \sim 10^{- 4}$.

As an additional outcome of this treatment, eq. (26) places an upper
limit to
the energy scale of inflation of
$M < m_{pl}(\delta \rho/\rho)^{2/3} \sim m_{pl}\cdot 10^{- 3}$,
above which thermally induced fluctuations would
be inconsistent with the observed density perturbations. Therefore, one
can also conclude that for thermally caused initial perturbations,
inflation should not occur earlier than about $m_{pl}/10^2$.

In addition to the amplitude fluctuation for the scalar mode, which was
treated in this paper, there is also an amplitude fluctuation for the
tensor mode \cite{star2}.  Since this involves weakly interacting gravitons,
a thermal mechanism for inducing these fluctuations seems less likely
to the standard treatment which considers quantum fluctuations.
\bigskip

Financial support was provided by the NSF INT-9301805 grant
and the U. S. Department of Energy, Division of High Energy and Nuclear
Physics.

\end{document}